\documentclass[aps,prl,twocolumn,longbibliography,superscriptaddress,footinbib,showpacs]{revtex4-1}
\usepackage[version=3]{mhchem}
\usepackage{amsmath,amssymb,graphicx}
\usepackage{bm}
\usepackage{natbib}
\usepackage{hyperref}

\hypersetup{pdfstartview={FitH},pdfpagemode={UseNone},
            colorlinks,linkcolor=blue, citecolor=blue, urlcolor=blue,
            bookmarks=true, bookmarksopen=true, pdfnewwindow=true}

\newcommand{\eps}{\varepsilon}
\newcommand{\rmi}{{\rm i}}
\newcommand{\rmd}{{\rm d}}
\newcommand{\rot}{\mathop{\mathrm{rot}}\nolimits}
\newcommand{\e}{{\rm e}}

\renewcommand{\Im}{\mathop{\mathrm{Im}}\nolimits}
\renewcommand{\Re}{\mathop{\mathrm{Re}}\nolimits}
\newcommand\REMOVE[1]{}

\begin{document}

\title{Generation of photon-plasmon quantum states \\in nonlinear hyperbolic metamaterials}

\author{\firstname{Alexander~N.} \surname{Poddubny}}
\affiliation{ITMO University, St. Petersburg 197101, Russia}
\affiliation{Ioffe Institute, St. Petersburg 194021, Russia}
\affiliation{Nonlinear Physics Centre, Research School of Physics and Engineering, Australian National University, Canberra, ACT 2601, Australia}

\author{\firstname{Ivan~V.} \surname{Iorsh}}
\affiliation{ITMO University, St. Petersburg 197101, Russia}

\author{\firstname{Andrey~A.} \surname{Sukhorukov}}
\affiliation{Nonlinear Physics Centre, Research School of Physics and Engineering, Australian National University, Canberra, ACT 2601, Australia}

\date{\today}

\pacs{42.65.Lm,78.67.Pt,42.82.Fv}
\email{a.poddubny@phoi.ifmo.ru}

\begin{abstract}
We develop a general theoretical framework of integrated paired photon-plasmon generation through spontaneous wave mixing in nonlinear plasmonic and metamaterial nanostructures, rigorously accounting for material dispersion and losses in quantum regime through the electromagnetic Green function.
We identify photon-plasmon correlations in layered metal-dielectric structures with 70$\%$ internal heralding quantum efficiency, and reveal novel mechanism of broadband generation enhancement due to topological transition in hyperbolic metamaterials.
\end{abstract}

\maketitle

Recent pioneering experiments demonstrated the quantum interference between individual photons in nanoscale plasmonic waveguides~\cite{Heeres:2013-719:NNANO}, operating up to the room temperature~\cite{Fakonas:2014-317:NPHOT, Fakonas:2015-23002:NJP}.
However, the photon generation relied on spontaneous wave mixing in external bulk nonlinear crystals. Further efforts are focused on the incorporation of photon sources in plasmonic and metamaterial structures, which on the one hand can lead to the realization of fully integrated quantum devices, and on the other hand can open new opportunities for manipulating the quantum features of emitted photons, for example through hyperbolic dispersion~\cite{Poddubny:2013-948:NPHOT, Ferrari:2015-1:PQE}.

The integrated photon-plasmon generation has been so far reported from quantum dot~\cite{Dousse:2010-217:NAT} and quantum well~\cite{Nevet:2010-1848:NANL} structures suffering from inhomogeneous broadening and dephasing.
A promising alternative approach to achieve coherent photon generation at room temperatures is to employ spontaneous nonlinear wave mixing processes, which are successfully used in conventional dielectric waveguide circuits~\cite{Silverstone:2014-104:NPHOT, Jin:2014-103601:PRL, Solntsev:2014-31007:PRX}. This route is feasible since plasmonic structures and metamaterials can enhance and precisely tailor nonlinear wave mixing~\cite{Popov:2006-131:APB, Suchowski:2013-1223:SCI, deCeglia:2014-75123:PRB, Duncan:2015-8983:SRP}.

To fully unlock the potential of the nanoscale plasmonic and metamaterial circuitry for integrated quantum state generation through spontaneous wave mixing, it is necessary to accurately model quantum nonlinear interactions in metal-dielectric structures, providing the fundamentals for structure design and simulation of experimental performance.
However the majority of theoretical techniques have been developed for conventional waveguide structures under the conditions of loss-less \cite{Yang:2008-33808:PRA} and nondispersive elements or including just a few optical modes~\cite{Drummond:2014:QuantumTheory, Crosse:2011-23815:PRA, Dezfouli:2014-43832:PRA, Antonosyan:2014-43845:PRA, Helt:2015-13055:NJP, Helt:2015-1460:OL, Onodera:1509.03180:ARXIV, Dezfouli:2015-JTu5A.12:ProcCLEO}.
Such methods are not suitable for plasmonic circuits, where frequency dispersion and metal losses are significant, and multiple spatial modes should be taken into account to describe photon emission~\cite{Poddubny:2013-948:NPHOT, Ferrari:2015-1:PQE}.

In this Letter we present a rigorous approach describing entangled photon-plasmon state generation through spontaneous wave mixing in metal-dielectric nanostructures of arbitrarily complex geometry. We derive ready-to-use explicit formulas for the experimentally measurable photon counts and quantum correlations. They are expressed through the classical electromagnetic Green function satisfying Maxwell equations and fully incorporating material absorption and dispersion characteristics.
We first demonstrate an application of our approach to a bilayer metal-dielectric structure and predict
photon-plasmon generation with $\gtrsim 70$~\% internal heralding efficiency. Then we analyze a multilayer metal-dielectric hyperbolic metamaterial~\cite{Poddubny:2013-948:NPHOT, Ferrari:2015-1:PQE}, where we reveal new type of photon-pair generation enhancement due to the broadband phase matching at the topological transition.

\REMOVE{
Q dots, atoms

two-photon emission

theory for localized source (atom), no detection consideration \cite{Poddubny:2013-948:NPHOT} - no spatial entanglement, but polarisation entanglement possible
"such a multiresonant
plasmonic environment engenders the possibility to achieve
multiparticle entanglement. This problem requires separate
studies since the entanglement depends not only on the
plasmonic environment, but also on the specific choice of the
quantum light source and on the excitation mechanism"
no correlations

experiments - purity not considered; measured correlations vs. delay \cite{NPHOT}

Reference to Cross and Scheel

Practical development of nonlinear plasmonic circuitry for quantum information applications requires integrated components for plasmon generation and manipulation.

Entangled states lie at the heart of quantum information, communications, and sensing devices~
}

\begin{figure}[b]
\begin{center}
\includegraphics[width=0.8\columnwidth]{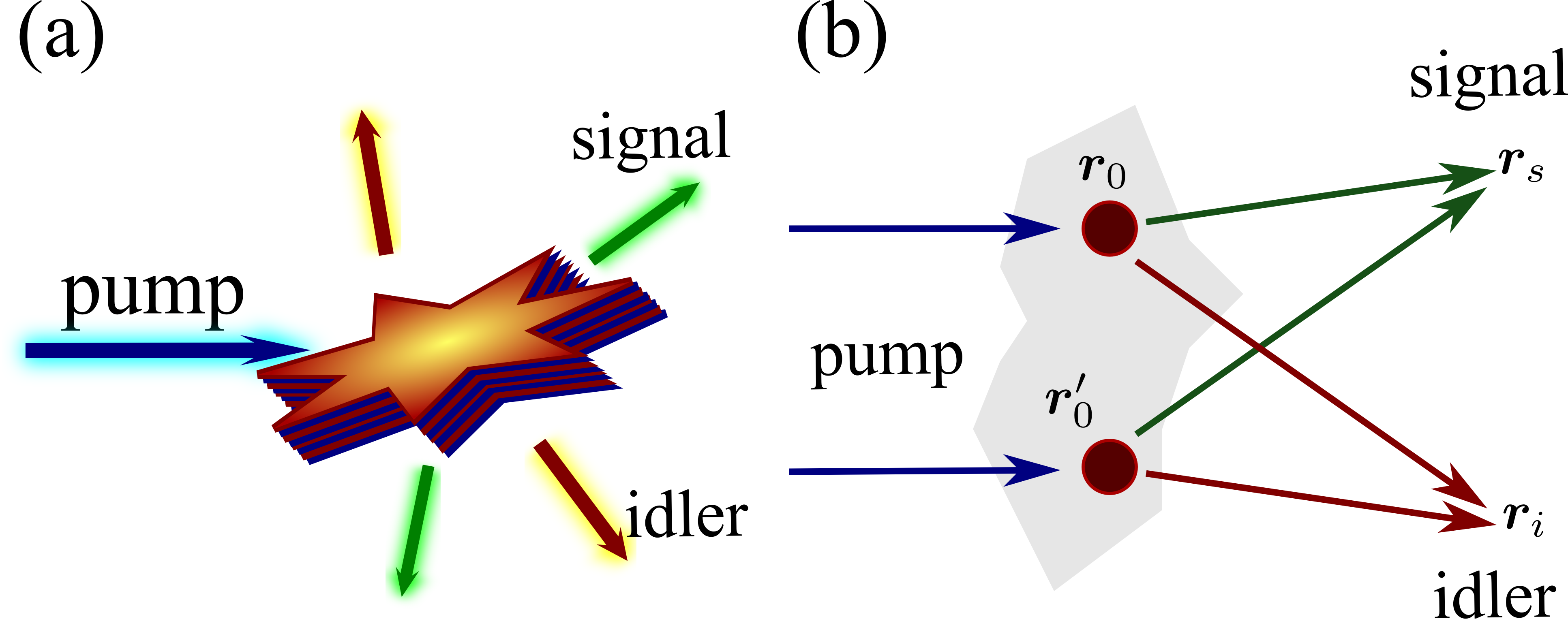}
\end{center}
\vspace*{-5mm}
\caption{\label{fig:1} (Color online)
(a) Scheme of the photon pair generation from a nonlinear metamaterial
(b)~Diagrammatic representation of the two-photon interference according to Eq.~\eqref{eq:Tis113}.
}
\end{figure}

We consider generation of a pair of signal and idler photons from a pump wave through spontaneous nonlinear wave mixing inside metal-dielectric structures, as schematically illustrated in Fig.~\ref{fig:1}(a).
The three relevant processes are the generation of the photon pairs, their (linear) propagation and possible absorption in the structure, and their detection.
Importantly, spontaneous nonlinear wave mixing realizes a spatially extended coherent source of photon pairs, which interference can lead to strong quantum entanglement even in presence of losses~\cite{Antonosyan:2014-43845:PRA, Helt:2015-13055:NJP, Helt:2015-1460:OL}. This is an important benefit compared to sets of quantum dots~\cite{Dousse:2010-217:NAT, Nevet:2010-1848:NANL} suffering from dephasing.
We consider the weak pumping regime neglecting the generation of multiple photon pairs. Then, in presence of linear losses, the output quantum state will be composed of pure photon pairs and single photons in a mixed state~\cite{Antonosyan:2014-43845:PRA, Helt:2015-13055:NJP, Helt:2015-1460:OL}.

The {\em photon-pair generation} is described by the Hamiltonian~\cite{Drummond:2014:QuantumTheory}
$ H_{\rm NL}={1}/{2}\int {\rmd \omega_{1}\rmd \omega_{2}}{(2\pi)^{-2}}
\rmd^{3}rE_{\alpha}^{\dag}(\omega_{1},\bm r)E_{\beta}^{\dag}(\omega_{2},\bm r)
\Gamma_{\alpha\beta}(\bm r)+\rm H.c.\: $,
where $E$ is the electric field operator, $\alpha,\beta=x,y,z$, and $\Gamma_{\alpha\beta}$ is the generation matrix.
We consider two possibilites~\cite{Boyd:2008:NonlinearOptics}, spontaneous parametric down conversion (SPDC) due to $\chi^{(2)}$ nonlinear susceptibility and spontaneous four-wave mixing (SFWM) governed by $\chi^{(3)}$ nonlinearity, when
\begin{equation} \nonumber
\Gamma_{\alpha\beta}(\bm r)=
\begin{cases}
\chi^{(2)}_{\alpha\beta\gamma}(\bm r; \omega_{1},\omega_{2};\omega_{p})\mathcal E_{p,\gamma}(\bm r) \e^{-\rmi\omega_{\rm p}t} ,\\*[9pt]
\chi^{(3)}_{\alpha\beta\gamma\delta}(\bm r; \omega_{1},\omega_{2};\omega_{p},\omega_{p})\mathcal E_{p,\gamma}(\bm r)\mathcal E_{p,\delta}(\bm r)\e^{-2\rmi\omega_{\rm p}t} ,\:
\end{cases}
\end{equation}
$\mathcal E_{p}$ is the classical pump at frequency $\omega_{\rm p}$,
$\gamma,\delta=x,y,z$.

The {\em linear propagation of the generated photons} is governed by the Hamiltonian
$
H_{\rm lin}=\int \rmd^{3}r\int_{0}^{\infty}\rmd \omega\hbar \omega \bm f^{\dag}\cdot \bm f
$,
where $f_{\alpha}(\bm r,\omega)$ are the canonical bosonic source operators for the quantum electric field~\cite{Vogel:2006:QuantumOptics}:
$\bm E(\bm r)=\int\limits_{0}^{\infty}{\rmd \omega}{(2\pi)^{-1}}\bm E(\bm r,\omega)+{\rm H.c.}$,
$\hat{\bm E}(\omega)= \rmi\sqrt{\hbar}\int \rmd^3\bm r'  G_{\alpha\beta}(\bm r,\bm r',\omega)\sqrt{\Im\eps(\omega,\bm r')}f_{\beta}(\bm r',\omega)$.
Here, $G$ is  the classical  electromagnetic Green tensor,
\begin{equation} \nonumber
 \left[\rot\rot-\left(\frac{\omega}{c}\right)^2\varepsilon(\omega,\bm r)\right]G(\bm r,\bm r',\omega)=4\pi \left(\frac{\omega}{c}\right)^2 \hat 1\delta(\bm r-\bm r')\:.
\end{equation}
The advantage of the local source quantization scheme~\cite{Vogel:2006:QuantumOptics} is the possibility to explicitly account for arbitrary strong Ohmic losses and mode dispersion, encoded in the Green function.
This method was previously applied~\cite{Poddubny:2012-33826:PRA} to describe the spontaneous two-photon emission (STPE)~\cite{Hayat:2008-238:NPHOT} from a single atom. However the current problem is quite distinct from STPE, because nonlinear spontaneous wave mixing acts as a coherent spatially extended source.

We explicitly introduce the {\em sensors that detect the quantum electromagnetic field}~\cite{delValle:2012-183601:PRL} to find the experimentally measurable quantities.
The sensors are modelled as  signal (s) and idler (i) two-level systems with the Hamiltonians
$
H_{i,s}=\hbar\omega_{i,s}a_{i,s}^{\dag}a_{i,s}-\hat{\bm d}_{i,s}\cdot \bm E(\bm r_{i,s})$,
with the resonant energies $\hbar\omega_{s}$ and $\hbar\omega_{i}$, respectively. Here,  $a^{\dag}_{s,i}$ are the corresponding exciton creation and
$\hat{\bm d}_{i,s}=a\bm d^{\ast}_{i,s}+a^{\dag}\bm d_{i,s}^{\ast}$ are the dipole momentum operators.

The detected two-quantum state is $|\Psi\rangle=a_{i}^{\dag}a_{s}^{\dag}|0\rangle$ with both detectors excited by the photon pair.
Formally, the process of photon pair generation, propagation, and detection can be described by the scattering matrix element $S_{is}=\langle \Psi|U|0\rangle$, where $U$ is the evolution operator~\cite{Cohen:1998:AtomPhoton}. We develop direct perturbation technique
(see Supplementary Materials) and obtain
$S_{is}=-2\pi \rmi \delta(\hbar\omega_{i}+\hbar\omega_{s}-N\hbar\omega_{\rm pump})T_{is}$, where $N=1(2)$ for SPDC (SWFM).
By construction the  two-photon transition amplitude  $T_{is}$ has the meaning of the complex wave function fully defining the pure two-photon state:
\begin{multline}
T_{is}(\bm r_{i},\omega_{i},\bm d_{i};\bm r_{s},\omega_{s},\bm d_{s})=\sum\limits_{\alpha\beta,\sigma_{i},\sigma_{s}}d^{\ast}_{i,\sigma_{i}}d^{\ast}_{s,\sigma_{s}}\\\times\int \rmd^{3} r_{0}
G_{\sigma_{i}\alpha}(\bm r_{i},\bm r_{0},\omega_{i})
G_{\sigma_{s}\beta}(\bm r_{s},\bm r_{0},\omega_{s})
\Gamma_{\alpha\beta}(\bm r_{0})\:.\label{eq:Tis113}
\end{multline}
This is the central result of our study. The form of Eq.~\eqref{eq:Tis113} clearly represents the interference between the spatially entangled photons generated in the different points of space $\bm r_{0}$~\cite{Ghosh:1987-1903:PRL}, as schematically illustrated in Fig.~\ref{fig:1}(b). 

The coincidence rate, which defines simultaneous detection of two photons at different positions in space, is found as:
$W_{is}=(2\pi/\hbar)\delta(\hbar \omega_{i}+\hbar\omega_{s}-N\hbar\omega_{p})|T_{is}|^{2}$. The single photon states can be measured by the total count rate of one detector, and for signal photons we obtain (see Supplementary Materials):
$W_{s}(\bm r_{s})= \frac{2}{\hbar}\iint \rmd^{3} r_{0}'\rmd^{3} r_{0}''\sum\limits_{\sigma_{s},\sigma_{s'}} d_{s,\sigma_{s}}d^\ast_{s,\sigma_{s'}}
\Im G_{\beta\beta'}(\bm r_{0}',\bm r_{0}'',\omega_{p}-\omega_{s})$
$\Gamma_{\alpha\beta}(\bm r_{0}')$ $\Gamma^{\ast}_{\alpha'\beta'}(\bm r_{0}'')$  $G_{\sigma_{s},\alpha}(\bm r_{s},\bm r_{0}',\omega_{s})$
$G_{\sigma_{s'},\alpha'}^{\ast}(\bm r_{s},\bm r_{0}'',\omega_{s})\:$.
We have verified that the general expressions exactly reduce to the previous results for waveguides~\cite{Antonosyan:2014-43845:PRA, Helt:2015-13055:NJP}, the details are given in the Supplementary Materials.

\begin{figure}[t]
\begin{center}
\includegraphics[width=0.8\columnwidth]{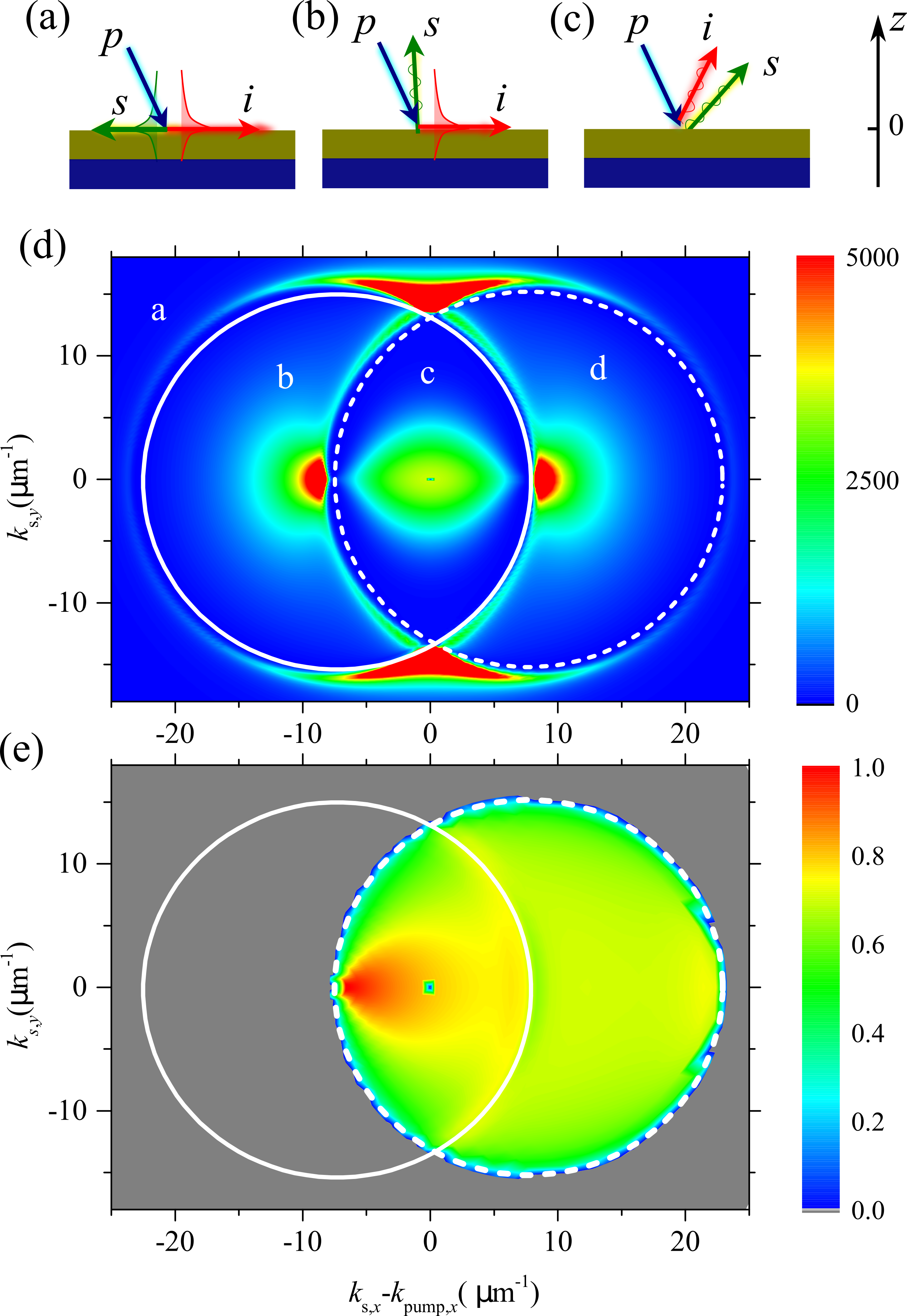}
\end{center}
\vspace*{-5mm}
\caption{\label{fig:2}
(Color online) (a)-(c) Scheme of SFWM generation of
a pair of (a) entangled plasmons, (b) entangled photons and (c) photon entangled with plasmon in the gold/nonlinear dielectric structure.
(d) Color map of the two-photon detection probability $|T(\bm k_{i},\bm k_{s})|^{2}$ in the reciprocal space vs. the in-plane wave vector components (arb.un.) in TM polarization ($\bm d_{i,s}\propto \bm k\times\hat{\bm z}$) at $z_{i}=z_{s}=100$~nm. The signal (solid) and idler (dashed) light lines are plotted in white color.
The letters a--d mark the near- and far-field signal and idler generation regimes.
(e) Efficiency of signal heralding by far field idler photons, Eq.~\eqref{eq:QE}.
For all plots $\hbar\omega_{i}\approx \hbar\omega_{s}\approx \hbar\omega_{p}\approx 3~$eV, $\eps_{\rm diel}=2$, $d_{\rm silver}=20$~nm, pump is TM polarized, $k_{p,x}=0.5\omega_{p}/c$, and silver permittivity according to Ref.~\cite{Johnson:1972-4370:PRB}.
}
\end{figure}

\begin{figure}[t]
\begin{center}
\includegraphics[width=\columnwidth]{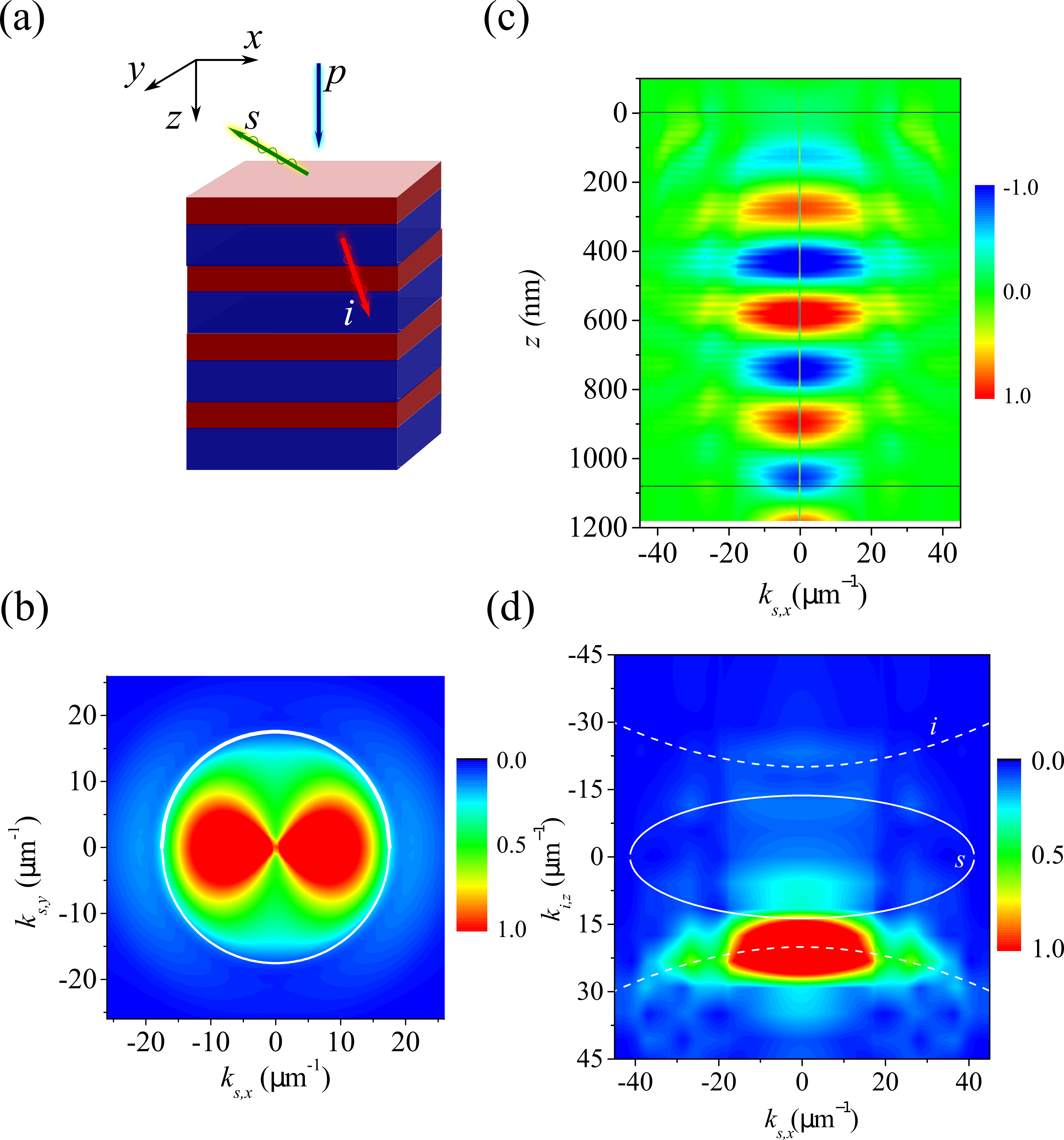}
\end{center}
\vspace*{-5mm}
\caption{ \label{fig:3}
(a)~Schematic illustration of two-photon generation from a nonlinear hyperbolic metamaterial.
(b)  Top view of  $|T_{is}(\bm k_{i},-\bm k_{s})|^{2}$ 
for $z=-100$~nm.
(c)  Side view of  $\Re T_{is}(k_{s,x},z_{i})$.
(d)  Side view of $|T_{is}(k_{x},k_{z})|^{2}$.
Calculated for TM polarizations of signal and idler detectors, normal pump incidence polarized along $x$, $\hbar\omega_{s}=3.46~$eV,
$\hbar\omega_{p}=3.6~$eV, $d_{\rm silver}=7.5$~nm, $d_{\rm diel}=15$~nm.
}
\end{figure}

We now apply the general theory to 
layered metal-dielectric plasmonic structures. First, we analyze the degenerate spontaneous four-wave mixing for the metallic layer on top of the nonlinear dielectric, see Figs.~\ref{fig:2}(a)--(c).
Due to the translational symmetry, the total in-plane momentum $\bm k$ of the photons and plasmons is conserved, i.e.
$
k_{i,\alpha}+k_{s,\alpha}=2k_{p,\alpha}$ for $\alpha=x,y$.
The most interesting situation is realized for oblique pump incidence, giving rise to four different regimes when (a) both signal and idler, (b) only idler, (c) neither signal nor idler and (d) only signal in-plane wave vectors lie outside the corresponding light cone boundaries $\omega_{i,s}/c$.
The first three situations are schematically shown in Figs.~\ref{fig:2}(a)--(c). Two-photon generation occurs in case (c), while (b) and (d) correspond to plasmon generation heralded by the far field photon.

We perform numerical simulations considering isotropic dielectric with {electronic} $\chi^{(3)}$ nonlinearity tensor as~\cite{Boyd:2008:NonlinearOptics}: $\chi_{\alpha\beta\gamma\delta}=\chi_{0}(\delta_{\alpha\beta}\delta_{\gamma\delta}
+\delta_{\alpha\delta}\delta_{\beta\delta}+\delta_{\alpha\gamma}\delta_{\beta\delta})$.
We plot the Fourier transform of the two-photon detection amplitude
$|T_{is}(\bm k_{s},z_{i},z_{s})|^{2}$ for $z_{i}=z_{s}=100~$nm above the structure, defined as
$T(\bm k_{i})=\int \rmd x\rmd y \exp(-\rmi k_{x}x-\rmi k_{y}y) T(x,y)$, which characterizes the signal-idler generation efficiency in all different regimes.
The relevant Fourier transforms of the Green functions were evaluated analytically following Ref.~\cite{Tomas:1995-2545:PRA}.
The overall map of the correlations resembles that for the generation of the polarization-entangled photons from a bulk nonlinear uniaxial crystal~\cite{Kwiat:1995-4337:PRL}: it shows strong maxima at the intersections of the signal and idler light cone boundaries. However, contrary to the bulk situation, the calculated map reflects the two-quantum correlations of both photons and plasmons.  In the region (c) the shown signal can be directly measured from the far field photon-photon correlations.  The near-field signal in the regions (a),(b),(d) can be recovered by using the grating to outcouple the plasmons to the far field~\cite{DiMartino:2012-2504:NANL} or with the near field scanning optical microscopy setup~\cite{leFeber:2014-43:NPHOT}. The optimization of the measurement  scheme for the specific sample can be handled by the presented general formalism, but it is out of the scope of the current study.

The bright spot in the map Fig.~\ref{fig:2}(d) for $k_{s,x}-k_{p,x}\approx 10~\mu$m$^{-1}$ reveals the resonantly enhanced plasmonic emission heralded by the normally propagating idler photons. The heralding efficiency can be estimated from the comparison of the signal-photon counts $W_{s}$ and the two-photon counts $W_{is}\propto |T_{is}|^{2}$. The details are given in the Supplementary  Information and the result reads
\begin{align}
QE&=\sum\limits_{z_{i}=-L,L}\sum\limits_{\bm d_{i}=\hat{\bm x},\hat{\bm y},\hat{\bm z}}\frac{c \cos\theta_{i}}{2\pi\hbar\omega_{i}}\frac{|T_{is}(\bm k_{s},z_{s},z_{i},\bm d_{i})|^{2}}{W_{s}(\bm k_{s})}\:,\label{eq:QE}
\end{align}
where
$\cos\theta_{i}=\sqrt{1-(ck_{i}/\omega_{i})^{2}}$.
The summation over $z_{i}$ in Eq.~\eqref{eq:QE} accounts for the total idler photon flux through the surfaces $z_{i}=\pm L$ above and below the nonlinear structure. The calculated values of the signal heralding, shown in Fig.~\ref{fig:2}(e), are remarkably high. They reach almost 100\% in the case when both signal and idler photon are in the far field, see the bright spot at $k_{s,x}-k_{p,x}\approx-5~\mu$m$^{-1}$. In the case of signal plasmons the heralding efficiency is uniform and about $70\%$.
We note, that the results in Fig.~\ref{fig:2}(e) correspond to the internal heralding efficiency, calculated for the plane pump wave.
The external quantum heralding efficiency has to account also for the plasmonic losses due to the propagation from the pump spot to the near field detector, which can be optimized in the actual experimental setup.

\begin{figure}[t]
\begin{center}
\includegraphics[width=\columnwidth]{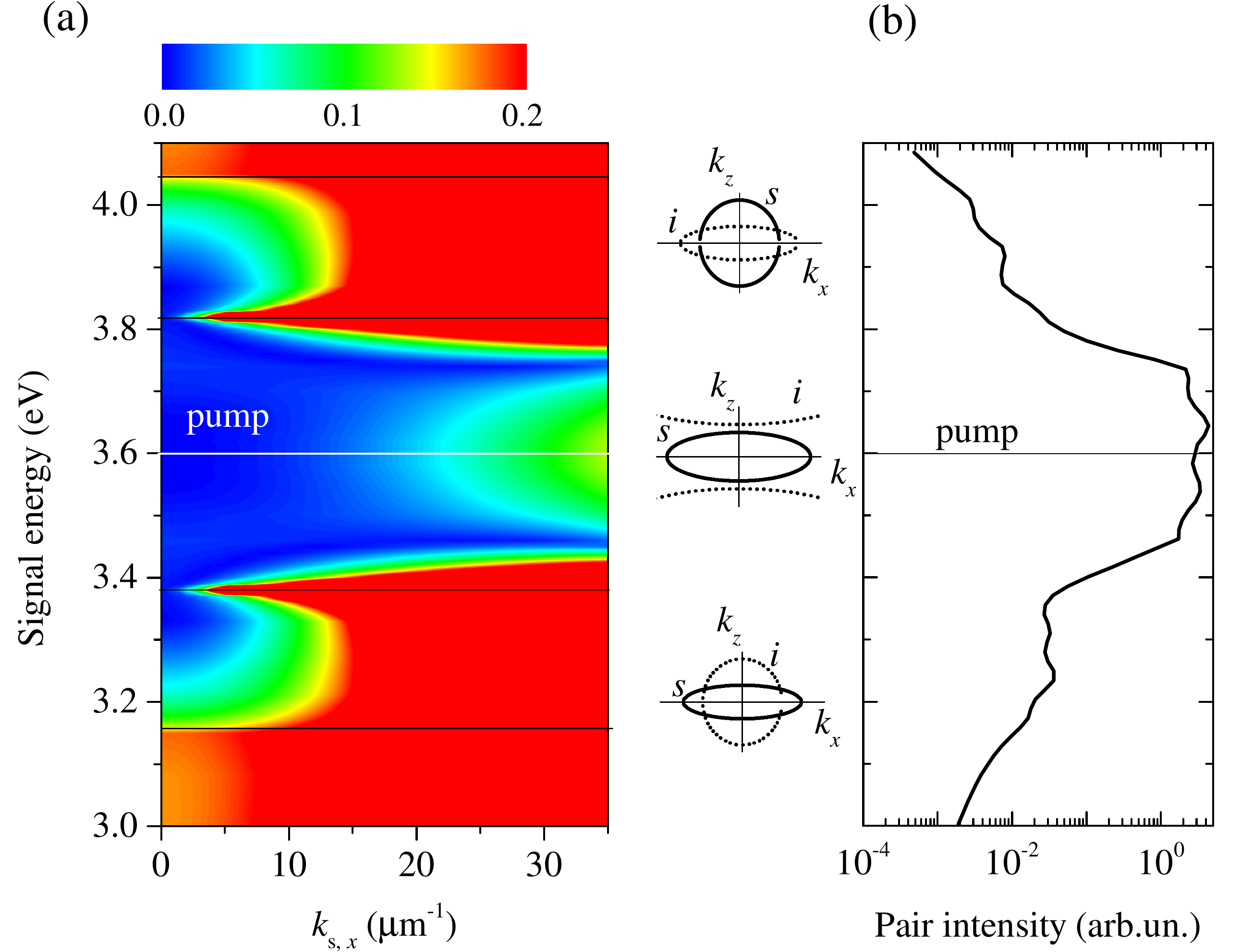}
\end{center}
\vspace*{-5mm}
\caption{(a) Phase matching map vs. the signal energy and in-plane wave vector. The color corresponds to $|\Re (k_{s,z}+k_{i,z}-2k_{pump,z})(d_{1}+d_{2})|$. Horizonal black lines show the boundaries between the spectral regions with topologically different dispersion of signal and idler photons (shown in the insets).
(b) Photon pair spectrum in the TM polarization for signal and idler, integrated over the wave vectors inside the signal light cone. Calculated for $\hbar\omega_{\rm pump}=3.6$~eV (indicated by a horizontal line) and the same other parameters as Fig.~\ref{fig:3}.
\label{fig:4}
}
\end{figure}

Next, we turn 
to the multilayered metal-dielectric hyperbolic metamaterial~\cite{Poddubny:2013-948:NPHOT, Ferrari:2015-1:PQE}. This is a strongly anisotropic artificial uniaxial medium, where the effective dielectric constant
$\eps_{xx}=\eps_{yy}$ and $\eps_{zz}$ can be of opposite signs, rendering the hyperbolic dispersion
law $k_{x}^{2}/\eps_{zz}+k_{z}^{2}/\eps_{xx}=(\omega/c)^{2}$ for the TM polarized waves.
We focus on the  non-degenerate SFWM.
Enhanced nonlinear processes such as Compton scattering \cite{Iorsh:2015-185501:PRL} and second harmonic generation
\cite{deCeglia:2014-75123:PRB, Duncan:2015-8983:SRP} have been recently predicted in the hyperbolic regime. The photon pair generation problem is quite different and remains open.

We consider  the pump  normally incident upon the metamaterial, see Fig.~\ref{fig:3}(a).
Generally, the enhanced local density of states in the hyperbolic metamaterials can not be harnessed  without the special outcoupling of the near field~\cite{Lu:2014-48:NNANO}. Here we avoid this obstacle by considering the non-degenerate spontaneous four-wave mixing  when the signal is in the elliptic regime and the idler is in the hyperbolic regime. This allows  the signal photons to escape the structure  while simultaneously making use of the enhanced density of states due to the hyperbolic plasmons at the idler frequency.

We present in Fig.~\ref{fig:3}(b) the two-photon correlations $|T_{is}(\bm k_{i},\bm k_{s})|^{2}$ in the reciprocal space in the $xy$ plane calculated for both signal and idler at $z_{i}=z_{s}=-100$~nm. The signal is concentrated in the far field region inside the light cone.
Figure~\ref{fig:3}(c) shows the side view  of the function $\Re T_{is}(k_{x},z)$ obtained for signal at $z_{s}=-100$~nm above the structure vs. the in-plane wave vector $k_{x}$ and idler detection coordinate $z_{s}$.
We observe the spatial oscillations of the pattern along the $z$ propagation direction for the hyperbolic idler plasmons within the metamaterial.
In order to get an insight of the plasmon propagation, we show in Fig.~\ref{fig:3}(d) the Fourier transform $|T_{is}(k_{x},k_{i,z})|^{2}$ as function of the idler photon wave vector. The solid and dashed white lines 
show the isofrequency contours at signal and idler frequencies. The maximum  of the two-photon response is pinned to the area between the elliptic signal dispersion and the hyperbolic idler dispersion.

To better understand the origin of the enhancement we analyze in Fig.~\ref{fig:4}(a) the phase matching conditions in the hyperbolic metamaterial by plotting the map of the momentum mismatch
$
|\Re (k_{i,z}+k_{s,z}-2k_{{\rm pump},z})|
$
vs. the signal energy and in-plane wave-vector. The map is symmetric with respect to the pump energy $\hbar\omega_{p}=3.6$~eV (white horizontal line), corresponding to degenerate SFWM. The phase mismatch exhibits dramatic changes when either signal or idler undergoes topological transition~\cite{Krishnamoorthy:2012-205:SCI} from the elliptic regime to the hyperbolic one (black lines).
The intermediate area for $3.4~{\rm eV}\lesssim\hbar\omega_{s}\lesssim 3.8$~eV corresponds to the phase matching realized in the broad band of in-plane wave vectors $k_{x}$  and frequencies $k_{z}$. The origin of the broad band phase matching is that the curvatures of the isofrequency contours $\rmd^{2}k_{z}/\rmd k_{z}^{2}$ are of opposite signs at the different sides of the topological transition where $\eps_{zz}$ changes sign. As a result, the contributions of signal and idler waves to the phase mismatch cancel each other.
Fig.~\ref{fig:4}(d) shows the spectrum of the integrated two-photon response [Fig.~\ref{fig:3}(c)] over the in-plane wave vector as function of the signal photon frequency. We observe a broad-band increase in the
spectral range $3.4~{\rm eV}\ldots 3.8$~eV when the phase matching is realized.

Finally, we note that our general result Eq.~\eqref{eq:Tis113} reveals an important quantum-classical correspondence for arbitrary structures with quadratic nonlinearity between the photon-pair generation through SPDC and sum-frequency generation (SFG) with classical signal and idler waves, which propagate in the opposite direction to the emulated signal and idler photons.
Namely, the far-field sum frequency signal $E_{\rm NL,\gamma}(\bm k_{NL},\omega_{i}+\omega_{s})$ is linked to the incident plane waves $\bm E_{s}\e^{\rmi \bm k_{s}\bm r-\rmi\omega_{s}t}$,
$\bm E_{i}\e^{\rmi \bm k_{i}\bm r-\rmi\omega_{i}t}$ as $E_{\rm NL,\gamma}=E_{s,\alpha}E_{i,\beta}T(\alpha,-\bm k_{i};\beta,-\bm k_{s})$, where $T(\alpha,-\bm k_{i};\beta,-\bm k_{s})$ is the Fourier component of Eq.~\eqref{eq:Tis113} evaluated for $\bm d_{i}=\bm e_{\alpha}$, $\bm d_{s}=\bm e_{\beta}$ and $\bm E_{p}=\bm e_{\gamma}\e^{-\rmi \bm k_{NL}\bm r-\rmi (\omega_{i}+\omega_{s})t}$. The direction-reversal was not considered previously, as only homogeneous lossy waveguides were analyzed~\cite{Helt:2015-1460:OL}. Due to the Lorentz reciprocity, the correspondence can be generalized to arbitrary reciprocal waves. This result will be reported in detail separately.

In conclusion, we developed a general theory for generation of photon and plasmon quantum pairs through spontaneous nonlinear wave mixing, applicable to any structure geometry and accounting for material dispersion and losses through the electromagnetic Green function.
We further predicted high internal heralding quantum efficiency and revealed topologically-enhanced phase-matching in layered metal-dielectric structures. This indicates the experimental feasibility in presence of metallic losses, and suggests even higher performance for all-dielectric nonlinear metamaterials and metasurfaces~\cite{Shcherbakov:2014-6488:NANL, Makarov:2015-6187:NANL, Jacob:2016-23:NNANO}. Moreover, our results can extend to other fields, including spontaneous four-wave-mixing in Bose-Einstein condensates loaded in tailored potentials~\cite{Lewis-Swan:2014-3752:NCOM}.

\begin{acknowledgments}
We acknowledge stimulating discussions with A.~V. Poshakinskiy, E.~L. Ivchenko, S. Saravi, M.~M. Glazov, A.~S. Solntsev, M. Steel.
This work was supported by the Australian Research Council (Discovery Project DP160100619), Russian Foundation for Basic Research and the ``Dynasty'' foundation.
\end{acknowledgments}

%

\end{document}